# QUANTUM MECHANICS OF SEMICONDUCTOR QUANTUM DOTS AND RINGS[1]

I. FILIKHIN, S.G. MATINYAN AND B. VLAHOVIC

*North Carolina Central University, 1801 Fayetteville St. Durham, NC 27707, USA*

We consider the several phenomena which are taking place in Quantum Dots (QD) and Quantum Rings (QR): The connection of the Quantum Chaos (QC) with the reflection symmetry of the QD, Disappearance of the QC in the tunnel coupled chaotic QD, electron localization and transition between Double Concentric QR in the transverse magnetic field ,transition of electron from QR to the QD located in the center of QR. Basis of this consideration is the effective Schrödinger equation for the corresponding systems.

## 1. Introduction

The progress of semiconductor physics in the decade 1970-1980 is connected with gradual deviation from the electronic band structure of ideal crystal of Bloch picture [1] where, unlike atomic world with its discrete and precisely defined, in the limits of uncertainty relation, energy levels, energy of bound electron is a multivalued function of momentum in the energy band and density of states are continuous (For the earlier short but comprehensive survey see [2]).

In principle, Bloch theory deals with infinite extension of lattice, with the understandable (and important) surface effects. The decreasing of the size of the object to a few micrometers principally does not change the picture of the extended crystal qualitatively. It takes a place until one reaches the scale where the size quantization essentially enters the game and we can speak about microscopic limit of matter. What generally divides macroscopic limit of the solid state from the microscopic one? It is defined by some correlation length (or, more generally, all such relevant lengths)): for carriers it is mean free path length l or Broglie length $l_B = h/p$ ( $p$ -momentum), which is smaller. One may say that the quantum mechanical properties of matter clearly reveal if $l/a \geq 1$, where $a$ is the size of the lattice constant. In the opposite limit $l/a < 1$, matter is considered macroscopically.

In this light, it is worthy to remind that as long as 1962, L. V. Keldysh [3] ([3] as cited in Ref. [4]) considered electron motion in a crystal with periodic potential with the period that is much larger than the lattice constant. In this limit he discovered so called minizones and negative resistance. Just in this limit $l/a \geq 1$ we expect the size quantization with its discrete levels and coherence in the sense that electron can propagate across the whole system without scattering, its wave function maintains a definite phase. In this limit, mesoscopic relates to the intermediate scale dividing the macro and micro limits of matter and nanoscopic objects (Quantum Wells (QW), Wires and Dots (QD)) shown very interesting quantum mechanical effects. In this limit many usual rules of macroscopic physics may not hold. For only one example, rules of addition of resistance both in series and parallel are quite different and more complicated [5-7].

---

[1] This paper is the basis of two talks of Sergei Matinyan ("Quantum Chaos and its disappearance in the Coupled Quantum Dots" (Nor-Amberd, 21 September 2011) and "Quantum Mechanics of Electron Transition between concentric Quantum Ring" (Tbilisi, 28 September 2011) at the workshop "Low Dimension Systems & Gauge principle" (Yerevan, Tbilisi).

Closing this brief introduction concerning some aspects of genuine quantum objects (QW, QWires, QD) we would like to emphasize the conditional sense of the notion of dimensions in this world: in the limit $l/a \geq 1$ dimensions are defined as difference between real spatial dimension (in our world $D=3$) and numbers of the confined directions: Quantum Well: $D=2$, Quantum Wire: $D=1$, Quantum Dot: $D=0$. However, for example, QD which will be one of our subject for study, has very rich structure with many discrete levels, their structure define the presence or absence of Chaos, as we will see below, inside QD. Minimal size of QD is defined by the condition to have at least one energy level of electron (hole) or both: $a_{min} = \pi \hbar / 2m^* \Delta E \sim 4$ nm, where $\Delta E$ is average distance between neighboring energy levels. Maximal size of QD is defined by the conditions that all three dimensions are still confined. It depends, of course, on temperature: at room temperature it is 12 nm (GaAs), 20 nm (InAs) ($\Delta E \approx 3kT$). The lower temperature, the wider QD is left as quantum object with $D=0$ and the number of energy levels will be higher.

## 2. Schrödinger equation and effective mass approximation

In the present review a semiconductor 3D heterostructure (QD or QR) is modelled utilizing a ***kp***-perturbation single sub-band approach with quasi-particle effective mass [8-10]. The energies and wave functions of a single carrier in a semiconductor structure are solutions the Schrödinger equation:

$$(H_{kp} + V_c(\vec{r}))\Psi(\vec{r}) = E\Psi(\vec{r}) \tag{1}$$

Here $H_{kp}$ is the single band ***kp***-Hamiltonian operator, $H_{kp} = -\nabla \frac{\hbar^2}{2m^*(\vec{r})} \nabla$, $m^*$ is the electron/hole effective mass for the bulk, which may depend on coordinate, and $V_c(\vec{r})$ is the confinement potential. The confinement of the single carrier is formed by the energy misalignment of the conduction (valence) band edges of the QD material (index 1) and the substrate material (index 2) in the bulk. $V_c(\vec{r})$ is so called "band gap potential". The magnitude of the potential is proportional to the energy misalignment. The band structure of the single band approximation one can be found in many textbooks (see, for example, [8-10]). We consider here the model in which the band gap potential is defined as follows:

$$V_c(\vec{r}) = \begin{cases} 0, & \in QD, \\ E_c, \vec{r} \notin QD, \end{cases}$$

where $E_c = \kappa(E_{g,2} - E_{g,1})$, $E_g$ is the band gap and the coefficient $\kappa < 1$ can be different for the conduction and valence bands gap potential. The BenDaniel-Duke boundary conditions are used on interface of the materials [11]: The single electron

Schrödinger equation for wave function $\Psi(\vec{r})$ and its derivative $1/m^*(\vec{n},\nabla)\Psi(\vec{r})$ on interface of QD and the substrate are continues.

### 3. The non-parabolicity of the conduction band. The Kane formula

Traditionally applied in the macroscopic scale studies parabolic electron spectrum needs to be replaced by the non-parabolic approach, which is more appropriate to nano-sized quantum objects [12, 13]. The Kane formula [14] is implemented in the model to take into account the non-parabolicity of the conduction band. The energy dependence of the electron effective mass is defined by the following formula:

$$\frac{m_0}{m^*} = \frac{2m_0 P^2}{3\hbar^2}\left(\frac{2}{E_g + E} + \frac{1}{E_g + \Delta + E}\right). \qquad (2)$$

Here $m_0$ is free electron mass, $P$ is Kane's momentum matrix element, $E_g$ is the band gap, and $\Delta$ is the spin-orbit splitting of the valence band.

Taking into account the relation (2) the Schrödinger equation (1) is expressed as follows

$$(H_{kp}(E) + V_c(\vec{r}))\Psi(\vec{r}) = E\Psi(\vec{r}). \qquad (3)$$

Here $H_{kp}(E)$ is the single band kp-Hamiltonian operator $H_{kp}(E) = -\nabla\frac{\hbar^2}{2m^*(E,\vec{r})}\nabla$, $m^*(E,\vec{r})$ is the electron (or hole) effective mass, and $V_c(\vec{r})$ is the band gap potential. As a result, we obtain a non-linear eigenvalue problem.

Solution of the problem (3)-(2) results that the electron/hole effective mass in QD (or QR) varies between the bulk values for effective mass of the QD and substrate materials. The same is given for the effective mass of carriers in the substrate. The energy of confinement states of carries is rearranged by the magnitude of the band gap potential $V_c$.

The Schrödinger equation (1) with the energy dependence of effective mass can be solved by the iteration procedure [15, 16, 17, 18].

$$\begin{aligned}H_{kp}(m^{*k-1}_i)\Psi^k(\vec{r}) &= E^k\Psi^k(\vec{r}),\\ m^{*k}_i &= f_i(E^k),\end{aligned} \qquad (4)$$

where $k$ is the iteration number, $i$ refers to the subdomain of the system; $i = 1$ for the QD, $i = 2$ for the substrate. $H_{kp}(m^{*k}_i)$ is the Hamiltonian in which the effective mass does not depend on energy and is equal to the value of $m^{*k}_i$, $f_i$ is the function defined by the relation (2). For each step of the iterations the equation (1) is reduced to Schrödinger equation with the effective mass of the current step which does not depend on energy. At the beginning of iterations the bulk value of the effective mass is employed. Obtained eigenvalue problem can be solved numerically (by the finite element method, for example). After that, a new value for effective mass is taken by using Eq. (2) and procedure is repeated. The convergence of the effective mass during the procedure has a place after 3-5 steps. As an example, the typical convergences for election effective mass and confinement energy of single electron are displayed in Fig. 1 for the InAs/GaAs QR [18]. Description of other methods for the solution of the problem (3)-(2) can be found in [19].

Remarks: at the first, in the present review the consideration was restricted by the electron and heavy hole carriers, and, the second, the Coulomb interaction was excluded. Often the linear approximation for the function $m*_i/m_0 = f(E,r)$ is used. We also will be applied the linear fit in the present paper.

## 4. Effective approach for strained InAs/GaAs quantum structures: effective potential

Here we propose the effective potential method to calculate the properties of realistic semiconductor quantum dot/ring nanostructures with the explicit consideration of quantum dot size, shape, and material composition. The method is based on the single sub-band approach with the energy dependent electron effective mass (Eq. (2)). In this approach, the confined states of carriers are formed by the band gap offset potential. Additional effective potential is introduced to simulate the cumulative band gap deformations due to strain and piezoelectric effects inside the quantum dot nanostructure. The magnitude of the effective potential is selected in such a way that it reproduces experimental data for a given nanomaterial.

We rewrite the Schrödinger equation (3) in the following form:

$$(H_{kp}(E) + V_c(\vec{r}) + V_s(\vec{r}))\Psi(\vec{r}) = E\Psi(\vec{r}). \qquad (5)$$

Here $H_{kp}(E)$ is the single band kp-Hamiltonian operator $H_{kp}(E) = -\nabla \frac{\hbar^2}{2m^*(E,\vec{r})} \nabla$. As previously, $m^*(E,\vec{r})$ is the electron (or hole) effective mass, and $V_c(\vec{r})$ is the band gap potential, $V_s(\vec{r})$ is the effective potential. $V_c(\vec{r})$ is equal zero inside the QD and is equal to $V_c$ outside the QD, where $V_c$ is defined by the conduction band offset for the bulk. The effective potential $V_c(\vec{r})$ has an attractive character and acts inside the volume of the QD. This definition for the effective potential is schematically illustrated by Fig. 2 for the conduction band structure of InAs/GaAs QD. In the figure, the confinement potential of the simulation model with effective potential $V_s$ is denoted as "strained" [20]. The band gap potential for the conduction band (valence band) can be determin as $V_c$=0.594 eV ($V_c$=0.506 eV). The magnitude of the effective potential can be chosen to reproduce experimental data. For example, the magnitude of $V_s$ for the conduction (valence) band chosen in [21] is 0.21 eV (0.28 eV). This value was obtained to reproduce results of the 8-th band **kp**-calculations of [22] for InAs/GaAs QD. To reproduce the experimental data from [23], the $V_s$ value of 0.31 eV was used in [20] for the conduction band.

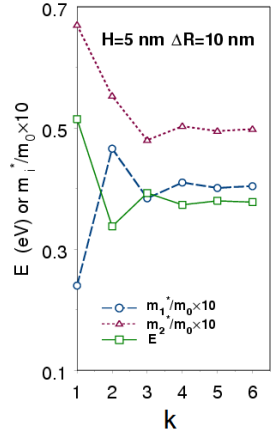

Figure 1. Convergence of the iterative procedure (4) for the confinement energy $E$ (solid line) and electron effective mass $m^*_i/m_0$ calculated for InAs/GaAs QR (dashed line) and GaAs substrate (dotted line). Here the height of QR is $H$, radial width is $\Delta R$ and inner radius is $R_1$ ($R_1$ =17 nm), $V_c$ =0.77 eV.

Possibility for the substitution of the function describing the strain distribution in QD and the substrate was firstly proposed in [24]. Recent works [25, 26] in which the strain effect taken into account rigorously applying the analytical method of continuum mechanics allow us to say that the approximation of the effective potential is appropriate.

In the next sub-section of the section 2 we will review the results obtained in both these approximations as the non-parabolic one as well as the effective potential method.

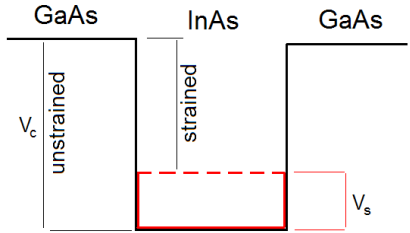

Figure 2. Effective potential $V_s$ and band gap structure of the conductive band of InAs/GaAs QD.

## 5. Electron energy in quantum rings with varieties of geometry: effect of non-parabolicity

In this section a model of the InAs/GaAs quantum ring with the energy dispersion defined by the Kane formula (2) (non-parabolic approximation) based on single sub-band approach is considered. This model leads to the confinement energy problem with three-dimensional Schrödinger equation in which electron effective mass depends on the electron energy. This problem can be solved using the iterative procedure (4). The ground state energy of confined electron was calculated of [17, 18, 27] where the effect of geometry on the electron confinement states of QR was studied and the non-parabolic contribution to the electron energy was estimated. The size dependence of the electron energy of QR and QD was subject of several theoretical studies [15, 28]. We present here, unlike the previous papers, a general relation for the size dependence of the QR energy.

Consider semiconductor quantum ring located on the substrate. Geometrical parameters of the semi-ellipsoidal shaped QR are the height $H$, radial width $\Delta R$ and inner radius $R_1$. It is assumed that $H/\Delta R \ll 1$ which is appropriate technologically. QR cross section is schematically shown in Fig. 3. The discontinuity of conduction band edge of the QR and the substrate forms a band gap potential, which leads to the confinement of electron.

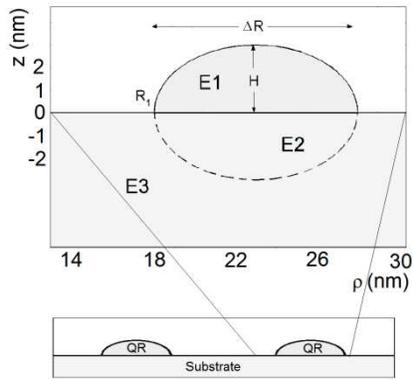

Figure 3. Profile of cross section of quantum ring (E1) and substrate (E2 and E3). Cylindrical coordinates $\rho$ and z shown on axis.

The band gap potential $V_c(\vec{r})$ is equal to zero inside the QR ($V_c(\vec{r})=0$) and it is equal to the confinement potential $E_c$ outside of the QR: The spatial dependence of the electron effective mass is given as $m^*(E,\vec{r}) = m_i^*(E)$, $i=1,2,3$, where $m_1^*$ is the effective mass in the material of QR ($\vec{r} \in$ E1), and $m_2^*(E)$, $m_3^*$ are the effective mass of the substrate material ($\vec{r} \in$ E2 and E3). Within each of the regions E1, E2 and E3 $m_i^*$ does not depend on the coordinates. The effective mass $m_3^*$ is equal to a constant bulk value. The energy dependence of the electron effective mass from the E1 and E2 subdomains is defined by the formula (2). The equation (1) satisfies the asymptotical boundary conditions: $\Psi(\vec{r})|_{|\vec{r}|\to\infty} \to 0$, $\vec{r} \in$ substrate and $\Psi(\vec{r})|_{\vec{r}\in S} = 0$, where $S$ is free surface of QR. On the surface of boundaries the wave function and the first order derivative $(\vec{n}, \vec{\nabla}\Psi)/m_i^*$ are continuous with different materials (the surface normal $\vec{n}$).

The Schrödinger equation (3) was numerically solved by the finite element method and iterative procedure (4). The following typical QR/substrate structures with experimental parameters were chosen: InAs/GaAs and CdTe/CdS. The parameters of the model are given in Tabl. 1 for the each hetero-structure.

Table 1: Parameters of the QR and substrate materials

| QR/Substrate | $m^*_1/m^*_2$ | $m^*_1/m^*_2$ | $2m_0P_1^2/\hbar^2 \,/\, 2m_0P_2^2/\hbar^2$ | $\Delta_1/\Delta_2$ |
|---|---|---|---|---|
| InAs/GaAs | 0.024/0.067 | 0.77 | 22.4/24.6 | 0.34/0.49 |
| CdTe/CdS | 0.11/0.20 | 0.66 | 15.8/12.0 | 0.80/0.07 |

It has to be noted that the effective mass substrate calculated for the InAs/GaAs and CdTe/CdS QRs is slightly differ from the bulk values within area E2. One can consider a simpler model when the properties of the area E2 and E3 are similar. It means that the wave function of electron does not penetrate through surface of QR (area E1) essentially. The simpler model does not change qualitative results of these calculations.

Analysis of the results of numerical calculations shows that the ground state energy of QR can be best approximated as a power function of the inverse values of the height and the radial width:

$$E \approx a(\Delta R)^{-\gamma} + bH^{-\beta}, \qquad (6)$$

where the coefficients $\gamma = 3/2$ and $\beta = 1$ were obtained numerically by the least square method. An example of this relation is illustrated in Fig. 4 for InAs/GaAs QR. Parameters a and b are remain constant except for extremely low values of $H$ and $\Delta R$. Our analysis also reveals a significant numerical difference between the energy of QR electron ground states, calculated in non-parabolic and parabolic approximations. The results of the calculation with parabolic approximation are represented in the Fig. 4 by the dashed lines.

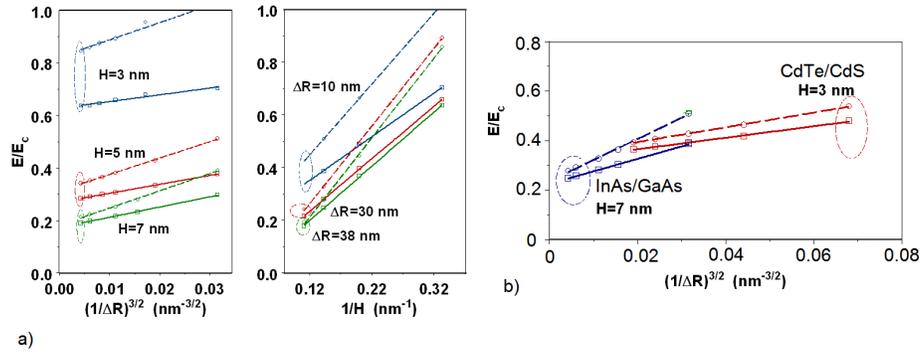

Figure 4. a) Normalized electron ground state energy of semi-ellipsoidal shape InAs/GaAs QR with parabolic (dashed line) and non-parabolic (solid lines) approximation as function of the QR size ($R_1 = 17$ nm). b) Normalized electron confinement energy of QRs of various materials in the parabolic (dashed line) and non-parabolic (solid lines) approximation.

Computation of the electron confinement energy of QRs for different materials show that the non-parabolic contribution is quite significant when chosen QR geometrical parameters are close to those of the QRs produced experimentally: $H < 7$ nm, $R < 30$ nm for InAs/GaAs, $H < 5$ nm, $R < 20$ nm for CdTe/CdS. Magnitude of this effect for InAs/GaAs can be greater than 30%. According with this fact, the coefficients $a$ and $b$ in Eq. (6) also depend on the approximation used: $a/b = 3.4/1.9$ for the non-parabolic and $a/b = 6.2/3.0$ for parabolic approximation.

As can be seen from the Fig. 4b), coefficients $\gamma$ and $\beta$ in the relation (6) do not depend on QR/substrate materials. Their values are defined by geometry and by the boundary conditions of the applied model. The model described above corresponds to the boundary condition as "hard wall at one side" (top side of the QR). For the model without the walls when the QR embedded into the substrate one can obtain $\gamma = 1$, and $\beta = 1/3$. In contrast with it, the coefficients $a$ and $b$ depend on the QR/substrate material set essentially.

Concluding, we have shown that for wide QR sizes the non-parabolicity effect does considerably alter the energy of the electron states, especially when the height or width of QR is relatively small.

## 6. The C-V measurements and the effective model: choosing the parameters

The well-established process of QDs formation by epitaxial growth and consecutive transformation of QDs into InAs/GaAs quantum rings (QR) [29] allows the production of 3D structures with a lateral size of about 40-60 nm and a height of 2-8 nm. In produced QDs and QRs it is possible directly to observe discrete energy spectra by applying capacitance-gate-voltage (CV) and far-infrared spectroscopy (FIR). In this section we will show how the effective model works using as an example the CV data. We use results of the CV experiment from [29, 30, 31] for QD and QR.

The effective mass of an electron in QD and QR changes from the initial bulk value to the value corresponding to the energy given by the Kane formula (2). Results of the effective model calculations for the InAs/GaAs QR are shown in Fig. 5. The effective mass of an electron in the InAs QR is close to that of the bulk value for the GaAs substrate. Since the effective mass in the QD is relatively smaller, as it is clear from Fig. 5, for QD the electron confinement is stronger; the $s$-shell peak of the CV trace is lower relative upper edge of conduction band of GaAs. The lower $s$-shell peak corresponds to the tunneling single electron into the QD. The picture is a starting point for the choosing the parameters of the effective potential model. In this section we follow the paper [20] where the semi-ellipsoidal InAs/GaAs QD has been considered. The average sizes of InAs/GaAs QD reported in Ref. [29] were: $H = 7$ nm (the height) and $R = 10$ nm (the radius). A cross section of the quantum dot is shown in Fig. 6a). The quantum dot has rotation symmetry. Thus the cylindrical coordinate was chosen in Eq. (5) which defines the effective model. For each step of iterative procedure (4) the problem (3-2) is reduced to a solution of the linear eigenvalue problem for the Schrödinger equation.

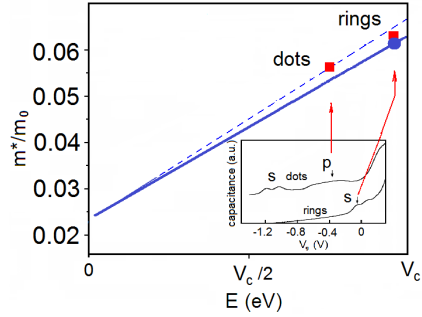

Figure 5. Calculated (circle) and experimentally obtained by [29,30] (squares) values for the electron effective mass and the confinement energies of the electron $s$- and $p$-levels of QD and QR. The solid line is obtained by the Kane formula (2), and the dashed line connects the bulk values of the effective mass. The inset: the capacitance-gate voltage traces [29].

Taking into account the axial symmetry of the quantum dot (ring) considered, this equation may be written in the cylindrical coordinates $(\rho, z, \phi)$ as follows:

$$(-\frac{\hbar^2}{2m^*}(\frac{\partial^2}{\partial \rho^2}+\frac{1}{\rho}\frac{\partial}{\partial \rho}-\frac{l^2}{\rho^2}+\frac{\partial^2}{\partial z^2})+V_c(\rho,z)+V_s(\rho,z)-E)\Phi(\rho,z)=0. \quad (7)$$

The wave function is of the form: $\Psi(r)=\Phi(\rho,z)exp(il\phi)$, where $l=0, \pm 1, \pm 2...$ is the electron orbital quantum number. For each value of the orbital quantum number $l$, the radial quantum numbers $n=0,1,2,...$ are defined corresponding to the numbers of the eigenvalues of (4) which are ordered in increasing. The effective mass $m^*$ must be the mass of electron for QD or for the substrate depending on the domain of the Eq. (2) is considered. The wave function $\Phi(\rho,z)$, and its first derivative in the form $\frac{\hbar^2}{2m^*}(\vec{n},\nabla)\Phi$, have to be continuous throughout the QD/substrate interface, where $\vec{n}$ is the normal vector to the interface curve. The Neumann boundary condition $\frac{\partial}{\partial \rho}\Phi(\rho,z)=0$ is established for $\rho=0$ (for case of QD). The asymptotical boundary conditions is $\Phi(\rho,z) \to 0$, when $\rho \to \infty$, $|z| \to \infty$ (QD is located near the origin of z-axes). When quantum dots are in an external perpendicular magnetic field, as it will be considered below, the magnetic potential term must be added to the potentials of Eq. (7) [32] in the form $V_m(\rho)=\frac{1}{2m^*}(\beta\hbar l+\frac{\beta^2}{4}\rho^2)$, where $\beta=eB$, $B$ is the magnetic field strength, and $e$ is the electron charge. We consider the case of a magnetic field normal to the plane of the QD and do not take into account the spin of electron because the observed Zeeman spin-splitting is small.

The confinement potential in Eq. (7) was defined as follow: $V_c$ = $0.7(E_g^S - E_g^{QD})$; $V_c = 0.77$ eV. The parameters of the QD and substrate materials were

$m^*_{bulk,1}/m^*_{bulk,2}$ =0.024/0.067, $\quad E_g^{QD}/E_g^S$ =0.42/1.52, $\quad \dfrac{2m_0 P_1^2}{\hbar^2}/\dfrac{2m_0 P_2^2}{\hbar^2}$ =20.5/24.6,

$\Delta_1/\Delta_2$ =0.34/0.49. The magnitude of the effective potential $V_s$ was chosen as 0.482 eV. There are three electron confinement states: the $s$, $p$, and $d$, as shown in the Fig. 5b). The energy of the $s$ single electron level measured from the top of the GaAs conduction band can be obtained from CV experimental data. To explain it, in Fig. 6c) the capacitance-gate-voltage trace from [33] is shown. The peaks correspond to the occupation of the $s$ and $p$ energy shells by tunneled electrons. The Coulomb interaction between electrons results to the $s$-shell splits into two levels and the $p$-shell splits into four levels if one takes into account the spin of electron and the Pauli blocking for fermions. The gate voltage-to-energy conversion coefficient $f=7$ ($\Delta E = e\Delta V_g / f$) was applied to recalculate the gate voltage to the electron energy. The value of the effective potential $V_s$ was chosen in order to accurately reproduce the observed $s$-wave level position with respect to the bottom of GaAs conduction band. The approximate size of this energy region is 180 meV.

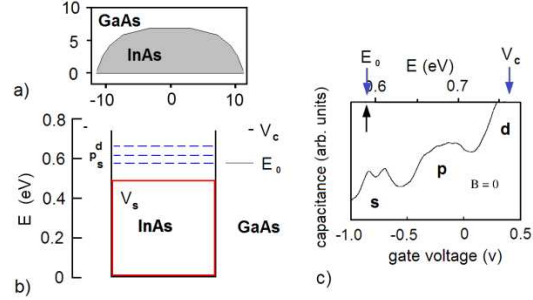

Figure 6. a) A cross section of the quantum dot. The dimensions are given in nm. b) Localization of the $s$, $p$ and $d$ single electron levels relatively to the bottom of the GaAs conduction band. $V_c$ is the band-gap potential, $V_s$ is the effective potential simulating the sum of the band-gap deformation potential, the strain-induced potential and the piezoelectric potential. c) The capacitance-gate-voltage trace [33]. The peaks correspond to the occupation of the $s$ and $p$ energy shells by tunneled electrons. The arrows denote the $s$ level ($E_0$) and the bottom of the GaAs conduction band.

The non-parabolic effect causes a change in the electron effective mass of QD with respect to the bulk value. According to the relation Eq. (2), the effective electron mass for InAs is sufficiently increased from the initial value of $0.024\,m_0$ to $0.054\,m_0$, whereas for GaAs substrate it is slightly decreased from $0.067\,m_0$ to $0.065\,m_0$ within the region of transmission of the wave function out of the quantum dot. The obtained value of the electron effective mass of InAs is close to the one ($0.057\,m_0 \pm 0.007$) extracted in Ref. [33] from the CV measurements of orbital Zeeman splitting of the $p$ level.

Applying the obtained effective model, one can take into account the effect the Coulomb interaction between electrons (the Coulomb blockade). The goal is to reproduce the C-V data presented in Fig. 6 for the InAs QD. The calculations [34] have been carried out using the perturbation procedure, proposed in [35]. The Coulomb energy matrix elements were calculated by applying single electron wave functions obtained from the numerical solution of Eq. (7). Both the direct terms of $E_{ij}^c$ and the exchange terms $E_{ij}^x$ of the Coulomb energy between electron orbitals with angular momentum projection of $\pm i$ and $\pm j$ were calculated (notation is given in [35]). The results of calculations of the electron energies of the $s$, $p$ and $d$ levels are shown in Fig. 3($Cal.$2). The $s$ shell Coulomb energy was found to be close to the experimental value which is about 20 meV.

Returning to the Fig. 5 we have to note that the effective potential obtained for InAs/GaAs QD has to be corrected for the case of the InAs/GaAs quantum rings. The reason is the topological, geometrical dependence of the depth of the effective potential. This dependence is weak for the considered QD and QR. The corresponding $V_s$ potentials have the magnitude of 0.482 eV and 0.55 eV for QD and QR, respectively. Accordingly to the experimental data the electron effective mass in quantum dots and rings changes from $0.024\,m_0$ to $(0.057\pm0.007)\,m_0$ [33] and $0.063\,m_0$ [29], respectively. The Kane's formula describes these variations well as it is shown in Fig. 5. The calculated values for the effective masses for quantum dots and rings are $0.0543\,m_0$ and $0.0615\,m_0$, respectively [34].

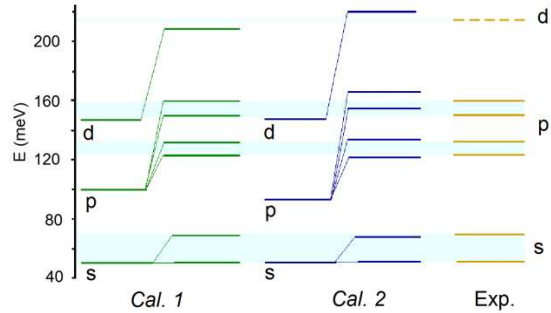

Figure 7. Energies of the electrons occupying a few first levels of the quantum dot at zero magnetic field. The calculations $Cal.$ 1 are that of parabolic model [35]. Our calculations are denoted by $Cal.$ 2. The splitting of the single electron levels of a corresponding energy shell is presented. CV experimental data are taken from [35].

Correct choice of the average QD profile is important for an analysis of the C-V data. It was shown in Ref. [36], where the calculation of the energy shifts due to the Coulomb interaction between electrons tunneling into the QD was performed for comparison with the C-V experiments.

One can see in Fig. 7 that the agreement between our results and the experimental data is satisfactory. Slight disagreement can be explained by uncertainty in the QD geometry which has not been excluded by available experimental data. In [36] it was shown that small variations of the QD cross section lead to significant changes in the levels presented in Fig. 7. The variations of the QD profile we considered are shown in Figure 8a, and the results of calculations for the electron energies are presented in Fig. 8b) for $s$ and $p$ – shell levels. The results of the calculations shown in Fig. 8 reveal rather high sensitivity to these variations of the QD profile. In particular, the spectral levels shift is noticeable due to a small deformation of the QD profile. Thus, we have seen that the average QD profile is important when we are comparing the result of the calculations and the experimental data.

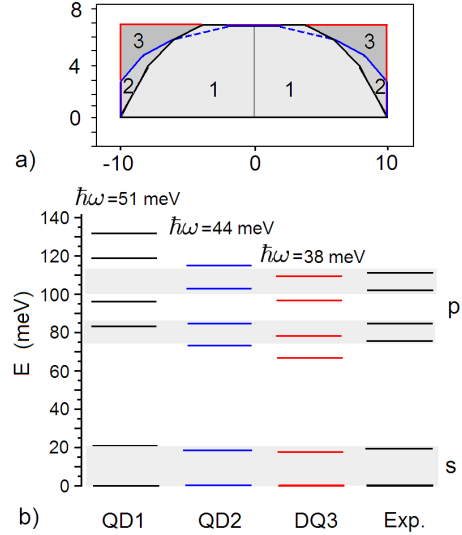

Figure 8. a) Cross sections of the QD. The dimensions are given in nm. b) Excitation energies of the electrons occupying $s$ and $p$ -energy shells of the InAs/GaAs quantum dot for various QD profiles are shown in Figure 7a). CV experimental data are taken from [35]. Here $\hbar\omega$ is the excitation energy $\hbar\omega = E_{(0,0)} - E_{(0,1)}$, where $E_{(n,l)}$ is a single electron energy of the $(n,l)$ state.

Finally, we may conclude that the effective model of QD/substrate semiconductor structure with the energy dependent effective mass and realistic 3D geometry taken into account, can quantitatively well interpret the CV spectroscopy measurements.

## 7. Experimental data for InAs/GaAs QR and the effective model

In this section we continue the description of the effective model use as an example the InAs/GaAs quantum ring. The geometry of the self-assembled QRs, reported in [29], is shown in Fig. 9 (Geometry 1). The InGaAs QRs have a height of about 2 nm, an outer diameter of about 49 nm, and an inner diameter of about 20 nm. Also, three-dimensional QR geometry (Geometry 2), which follows from the oscillator model [31] is used. The confinement of this model is given by the parabolic potential: $U(r) = \frac{1}{2}m^*\omega(r-r_0)^2$, where $\omega$, $r_0$ are parameters [37]. The QR geometry constructed by the relation between the adopted oscillator energy and a length $l$ as follows [38]:

$$l = \sqrt{2\hbar/m^*\omega}. \tag{8}$$

Here the width $d$ for the considered rings is defined by $d = 2l$. The obtained geometry with the parameters $m^*$ and $\omega$ from Ref. [31] is shown in Fig. 9 (Geometry 2); $m^* = 0.067\, m_0$ and $\omega = 15$ meV. The averaged radius of QR is 20 nm.

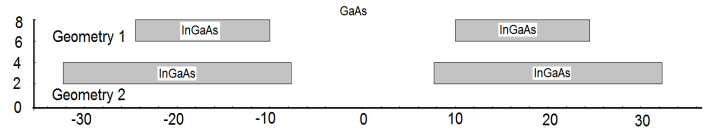

Figure 9. QR cross section profile corresponding to Geometry 1 and Geometry 2; sizes are in nm.

Results of the effective model calculations for the ground state energy of electron in a magnetic field are shown in Fig. 10 [39]. The picture of the change of the orbital quantum number of the ground state is similar to that obtained in ref. [31] with the oscillator model. The change occurred at 2.2 T and 6.7 T. The obtained energy fits the experimental data rather well.

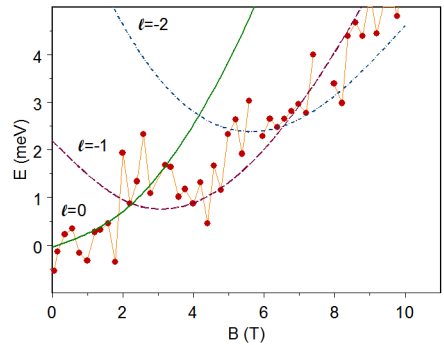

Figure 10. Additional energy of an electron in QR in a magnetic field $B$. The C-V experimental energies (circles) were obtained in Ref. [31] by using a linear approximation $\Delta E = e\Delta V_g / f$, with the lever arm $f = 7.84$. The curves $l = 0, -1, -2$ are the results of our calculations multiplied by a factor of 1.18 [31].

It has to be noted here that one cannot reproduce this result using the geometry proposed in Ref. [31] (Geometry 1) for this QR. The correspondence between the confinement potential parameters of the oscillator model and the real sizes of quantum objects has to be formalized by Eq. (8). Only using the geometry followed from Eq. (8) can we reproduce result of Ref. [31], as is shown in Fig. 10. The strength parameter of the effective potential, in the case of the Geometry 2, was chosen to be 0.382 eV, which is close to that for QD from Ref. [36], where $V_s = 0.31$ eV. The difference is explained by the topology dependence of the effective potentials (see section above and also [20]).

Note that the considered QRs are the plane quantum rings with the condition $H \ll D$, which enhances the role of the lateral size confinement effect. To qualitatively represent the situation shown in Fig. 12, one can used an approximation for the 3D QR based on the formalism of one dimensional ideal quantum ring. Additional electron energy, due to the magnetic field, can be calculated by the relation: $E = \hbar^2/(2m^*R^2)(l+\Phi/\Phi_0)^2$ (see for instance [30]), where $\Phi = \pi R^2 B$, $\Phi_0 = h/e$ ($\Phi_0 = 4135.7$ T nm2); $R$ is radius of the ideal ring. The Aharonov-Bohm (AB) [40] period $\Delta B$ [41] is estimated by the relation: $\Delta B = \Phi_0/\pi/R^2$. Using the root mean square (rms) radius for $R$ ($R = 20.5$ nm), one can obtain $\Delta B/2 = 1.56$ T and $\Delta B/2 + \Delta B = 4.68$ T for the ideal ring. This result is far from the result of 3D calculations shown in Fig. 12 where $\Delta B/2 \approx 2.2$ T and $\Delta B/2 + \Delta B \approx 6.7$ T are determined. Note here that the electron root mean square radius $R_{n,l}$ is defined by the relation: $R^2_{n,l} = \int |\Phi^N_{n,l}(\rho,z)|^2 \rho^3 d\rho dz$, where $\Phi^N_{n,l}(\rho,z)$ is the normalized wave function of electron state described by the quantum numbers ($n,l$).

One can obtain better agreement by using the radius for the most probable localization of the electron $R_{loc.}$, defined at the maximum of the square of the wave function. The electron is mostly localized near 17.1 nm, for $B = 0$. With this value, the ideal ring estimation leads to the values for $\Delta B/2$ and $\Delta B/2 + \Delta B$ as 2.25 T and 6.75 T, respectively. That agrees with the result of the 3D calculations (see Fig. 10). Obviously, the reason for this agreement is the condition $H \ll D$, for the considered QR geometry as it was mentioned above. The mostly localized position of the electron in QR depends weakly on the magnetic field. We present $R_{loc.}$ as a function of the magnetic field $B$ in Fig. 11. $R_{loc.}(B)$ is changed in an interval of $\pm 1$ nm around the mean value $R_{loc.}(0)$ of 17 nm. It is interesting to note that the magnetization of a single electron QR demonstrates the same behavior as it does for $R_{loc.}(B)$ if the one dimensional ring (circle) is used in Ref. [32] (see Ref. [32] for details).

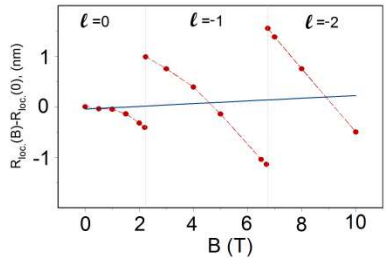

Figure 11. The radius ($R_{loc.}$) of the most localized position of an electron as a function of a magnetic field $B$. The electron of the ground state is considered. The circles indicate the calculated values and the solid line indicates the result of the least squares fitting of the calculated values. The orbital quantum number of the ground state is shown.

Additionally we compare the results of calculations for the QR geometry parameters corresponding to Geometry 1 and Geometry 2 in Fig. 9 with the far-infrared (FIR) data, reported in Ref. [35]. The results are presented in Fig. 12. One can see that the QR geometry proposed in Ref. [31] leads to a significant difference between the FIR data and the effective model calculations (see Fig. 12a), whereas the results obtained with Geometry 2 are in satisfactory agreement with the data (Fig. 12b). Again we conclude that the QR geometry of [31] does not provide an adequate description of electron properties of the InAs/GaAs QRs measured in [29, 31].

To summarize, we wish to point out that the problem of reliable theoretical interpretation of the C-V (and FIR) data for InAs/GaAs quantum rings is far from resolved. Obtained geometry can be considered as a possible version of geometry for experimentally fabricated QR.

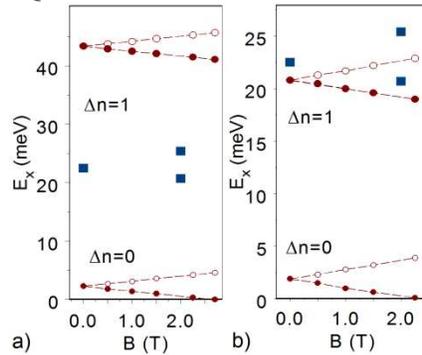

Figure 12. Solid squares represent the observed resonance positions [30] of the FIR transmissions at various magnetic field $B$. Calculated energies of the excited states with $|\Delta l|=1$ are marked by the circles. a) QR with shape given by Geometry 1, b) QR with shape given by Geometry 2. The orbital quantum number of the ground state is $l=0$. The quantum number $n$ is changed as it shown.

## 8. Quantum Chaos in Single Quantum Dots

### 8.1 Quantum Chaos

Quantum Chaos concerns with the behavior of quantum systems whose classical counterpart displays chaos. It is quantum manifestation of chaos of classical mechanics.
The problem of quantum chaos in meso - and nano-structures has a relatively long history just since these structures entered science and technology. The importance of this problem is related to wide spectrum of the transport phenomena and it was actively studied in the last two decades [42, 43, 44]. One of the main results of these studies, based mainly on the classical and semi-classical approaches, is that these phenomena sensitively depend on the geometry of these quantum objects and, first of all, on their symmetry: Right - Left (RL) mirror symmetry, up-down symmetry and preserving the loop orientation inversion symmetry important in the presence of the magnetic field [45, 46].

These results are well -known and discussed widely. There is another, actively studied in numerous fields of physics, aspect which ,in essence, is complimentary to the above mentioned semi classical investigations: Quantum Chaos with its inalienable quantum character, including, first of all, Nearest Neighbor level Statistics (NNS ) which is one of the standard quantum-chaos test.

Mathematical basis of the Quantum Chaos is a Random Matrix Theory (RMT ) developed by Wigner, Dyson, Mehta and Goudin (for comprehensive review see book

[42]). RMT shows that the level repulsion of quantum systems (expressed by one of the Wigner-Dyson -like distributions of RMT) corresponds to the chaotic behavior and, contrary, level attraction described by Poisson distribution tells about the absence of chaos in the classical counterpart of the quantum system. This theorem-like statement checked by numerous studies in many fields of science. For the completeness, we add that there are other tests of Quantum Chaos based on the properties of the level statistics: $\Delta_3$ statistics (spectral rigidity $\Delta_3(L)$), Number variance $\Sigma_2(L)$), spectral form-factor, two- and multipoint correlation functions, two level cluster function $Y_2(E)$ etc. They play an important subsidiary role to enhance and refine the conclusions emerging from the NNS.

The present review surveys the study of the NNS of nanosize quantum objects - quantum dots (QD) which demonstrate atom-like electronic structure under the regime of the size confinement. To use effectively NNS, we have to consider so called weak confinement regime where the number of levels can be of the order of several hundred. QD of various shape embedded into substrate are considered here under the effective model [47]. We use the sets of QD/substrate materials (Si/SiO$_2$, GaAs/Al0.7Ga0.25As, GaAs/InAs).

*8.2 The nearest neighbor spacing statistics*

For the weak confinement regime (for the Si/SiO$_2$ QD, the diameter $D \geq 10$ nm), when the number of confinement levels is of the order of several hundred [47], we studied NNS statistics of the electron spectrum. The low-lying single electron levels are marked by $E_i$, $i = 0,1,2,...N$. One can obtain the set $\Delta E_i = E_i - E_{i-1}$, $i = 1,2,3,...N$ of energy differences between neighboring levels. An example of the energy spectrum and set of the neighbor spacings for Si/SiO$_2$ QD are in Fig. 13. We need to evaluate the distribution function $R(\Delta E)$, distribution of the differences of the neighboring levels. The function is normalized by $\int R(\Delta E) d\Delta E = 1$. For numerical calculation, a finite-difference analog of the distribution function is defined by following relation:

$$R_j = N_j / H_{\Delta E} / N, \; j = 1,...M,$$

where $\sum N_j = N$ represents total number of levels considered, $H_{\Delta E} = ((\Delta E)_1 - (\Delta E)_N)/M$ is the energy interval which we obtained by dividing the total region of energy differences by $M$ bins. $N_j$ ( $j = 1,2,...M$ ) is the number of energy differences which are located in the $j$ -th bin.

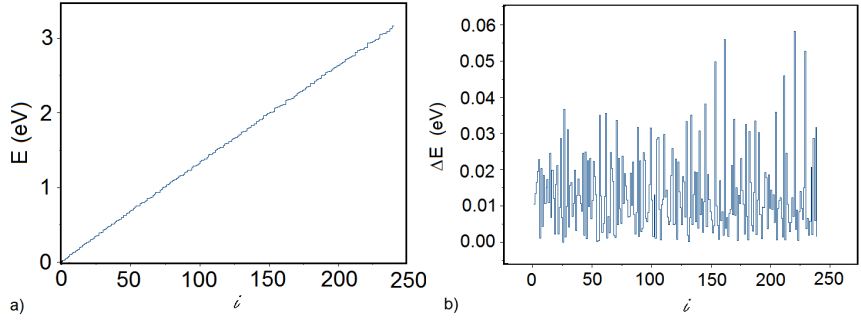

Fig. 13. a) The energy levels of the spherical Si/SiO2 QD with diameter $D$ =17 nm. The parameters $N$ =245, $M$ =9 were used. b) The neighbor spacings $\Delta E_i = E_i - E_{i-1}$, $i = 1,2,..N$ for of the spherical Si/SiO$_2$ QD with diameter $D$ =17 nm. The parameters $N$ =245, $M$ =9 were used.

The distribution function $R(\Delta E)$ is constructed using the smoothing spline method. If $R_j$, $j = 1,2,..M$ are calculated values of the distribution functions corresponding to $\Delta E_j$, the smoothing spline is constructed by giving the minimum of the form $\sum_{j=1}^{M}(R_j - R(\Delta E_j))^2 + \int R''(\Delta E)^2 d(\Delta E)/\lambda$. The parameter $\lambda > 0$ is controlling the concurrence between fidelity to the data and roughness of the function sought for. For $\lambda \to \infty$ one obtains an interpolating spline. For $\lambda \to 0$ one has a linear least squares approximation.

We studied neighbouring level statistics of the electron/hole spectrum treated by way described above. The Si quantum dots having strong difference of electron effective mass in two directions is considered as appropriate example for the study of role of the effective mass asymmetry. In this study we do not include the Coulomb potential between electrons and holes. The shape geometry role is studied for two and three dimensions.

*8.3 Violation of symmetry of the QD shape and nearest neighbor spacing statistics*

Distribution functions for the nearest neighboring levels are calculated for various QD shapes [47]. Our goal here to investigate the role of violation of the QD shape symmetries on the chaos. The two and three dimensional models are considered. It is important to note that existing of any above mentioned discrete symmetry of QD shape leads to the Poisson distribution of the electron levels.

In Fig. 14 the numerical results for the distribution functions of Si/SiO$_2$ QD are presented. The QD has three dimensional spherical shape. We considered the two versions of the shape. The first is fully symmetrical sphere, and the second shape is a sphere with the cavity damaged the QD shape. The cavity is represented by semispherical form; the axis of symmetry for this form does not coincide with the axis of symmetry of the QD. In the first case, the distribution function is the Poisson-like distribution. The violation symmetry in the second case leads to non-Poisson distribution.

We fit the distribution function $R(\Delta E)$ using the Brody distribution [48]:

$$R(s) = (1+\beta)bs^{\beta}\exp(-bs^{1+\beta}), \tag{10}$$

with the parameter $\beta = 1.0$ and $b = (\Gamma[(2+\beta)/(1+\beta)]/D)^{1+\beta}$, $D$ is the average level spacing. Note that for the Poisson distribution the Brody parameter is zero.

If the QD shape represents a figure of rotation (cylindrical, ellipsoidal and others) then the 3D Schrödinger equation is separable. In cylindrical coordinates the wave function is written by the following form $\psi(\vec{r}) = \Phi(\rho, z)\exp(il\varphi)$, where $l = 0, \pm 1, \pm 3,...$ is the electron orbital quantum number. The function $\Phi(\rho, z)$ is a solution of the two dimensional equation for cylindrical coordinates $\rho$ and $z$.

Our results for the distribution function for the ellipsoidal shaped Si/SiO$_2$ QD are presented in Fig. 15a) (left). In the inset we show the cross section of the QD. The fitting of the calculated values for $R(\Delta E)$ gives the Poisson-like distribution. For the case of QD shape with the break of the ellipsoidal symmetry (Fig. 15b) (left)) by the cut below the major axis we obtained a non-Poisson distribution.

Fig. 15 (right) It is shown that slightly deformed rhombus-like shape leads to the NNS with Brody parameter $\beta = 1$ (10). It is obvious why systems with different discrete symmetries reveal Poisson statistics: the different levels of the mixed symmetry classes of the spectrum of the quantum system are uncorrelated.

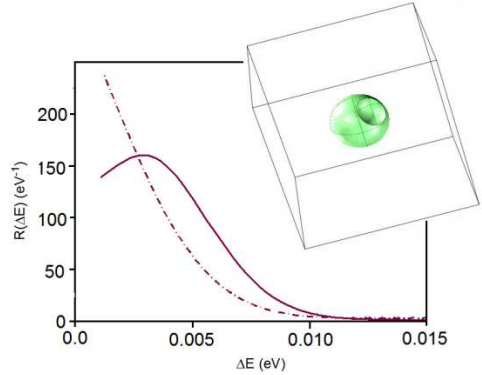

Fig. 14. Distribution functions for electron neighboring levels in Si/SiO2 QD for spherical-like shape with cut. The Brody parameter $\beta = 1.0$. The geometry of this QD is shown in 3D. The QD diameter is 17 nm (see the inset).

In Schrödinger equation (7) in the asymptotical region of large $\rho$ one can neglect the two terms $\frac{1}{\rho}\frac{\partial}{\partial \rho}$ and $-\frac{l^2}{\rho^2}$ of this equation. The solution of Eq. (7) can demonstrate the same properties of the solution of the Schrödinger equation for 2D planar problem in Cartesian coordinates with the same geometry of QD shape in the asymptotical region. We illustrate this fact by Fig. 16. In this figure the violation of the shape Up-Down symmetry for 2D Si/SiO2 QD is clarified. We compare the distribution functions for QD with "regular" semi-ellipsoidal shape (dashed curve in Fig. 16a) and for QD with the semi-ellipsoidal shape having the cut (solid curve) as it are shown in Fig. 16b). In the first case there is Up-Down symmetry of the QD shape. Corresponding distribution functions has the Poissonian type. In second case the symmetry is broken by cut. The level statistics become to non Poissonian. We have qualitative the same situation as for QD having rotation symmetry in 3D, presented in Fig. 15 (left) for the QD shape with rotation symmetry in cylindrical coordinates. The relation between the symmetry of QD shape and NNS is presented visually by Fig. 17 where we show the results of calculation of NNS for the 2D InAs/GaAs quantum well (QW). The two types of the statistics are presented in Fig. 20(left). The Poisonian distribution corresponds to shapes shown in Fig. 17 (b)-(d)(left) with different type of symmetry. The non-Poissonian distribution has been obtained for the QW shape with cut (a) which violates symmetry of initial shape (b), which is square having left-right symmetry, up-down symmetry, and diagonal reflection symmetry. The shape of the Fig. 17c) has only diagonal reflection symmetry. In Fig. 17d) the left-right symmetry of the shape exists only. The electron wave function of the high excited state, which contour plot is shown with the shape contour in Fig. 17(left), reflects the symmetry properties of the shapes.

## 9. Double Quantum Dots and Rings: new features

### 9.1 Disappearance of Quantum Chaos in Coupled Chaotic Quantum Dots

In the previous section, we investigated the NNS for various shape of the single quantum dots (SQD) in the regime of the weak confinement when the number of the levels allows to use quite sufficient statistics. Referring for details to [47], we briefly sum up the main conclusions of previous section: SQDs with at least one mirror (or rotation) symmetry have a Poisson type NNS whereas a violation of this symmetry leads to the Quantum Chaos type NNS.

In this section we study quantum chaotic properties of the double QD (DQD). By QD here we mean the three dimensionally (3D) confined quantum object, as well its 2D analogue - quantum well (QW). In three dimensional case we use an assumption of the rotational symmetry of QD shape. The presented effective approach is in good agreement with the experimental data and previous calculations in the strong confinement regime [47]. Here, in the regime of weak confinement, as in Ref. [47], we also do not consider Coulomb interaction between electron and hole: Coulomb effects are weak when the barrier between dots is thin leading to the strong interdot tunneling and dot sizes are large enough. In these circumstances, studied in detail in [49] (see also for short review a monograph [4], one may justify disregard of the Coulomb effects. The physical effect, we are looking for, has place just for thin barriers; to have sufficient level statistics, we need large enough QDs ( $\geq 100$ nm for InAs/GaAs QW).

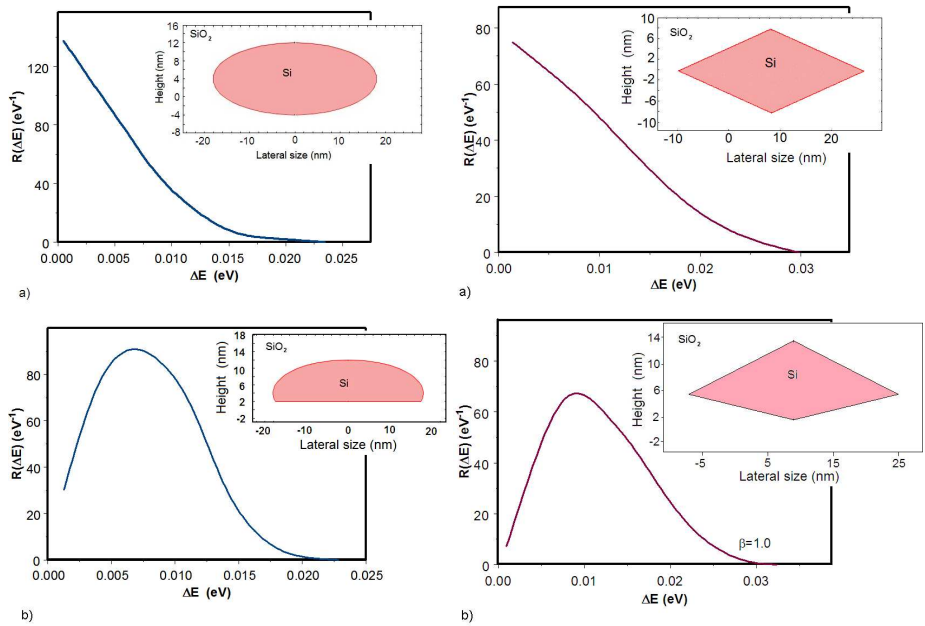

Figure 15 (Left) Distribution functions for electron neighboring levels in Si/SiO$_2$ QD for different shapes: a) ellipsoidal shape, b) ellipsoidal like shape with cut. Brody parameter $\beta$ is defined to be equal 1.02 for the fitting of this distribution. The 3D QD shape has rotation symmetry. Cross section of the shapes is shown in the inset.
(Right). Violation of the shape Up-Down symmetry for Si/SiO$_2$ QD. Distribution functions for electron neighboring levels in Si/SiO$_2$ QD for different shapes: a) with rhombus cross section, b) with slightly deformed rhombus cross section. The 3D QD shape has rotation symmetry. The Brody parameter $\beta$ for the curve fitting this distribution is shown. Cross section of the shapes is shown in the inset.

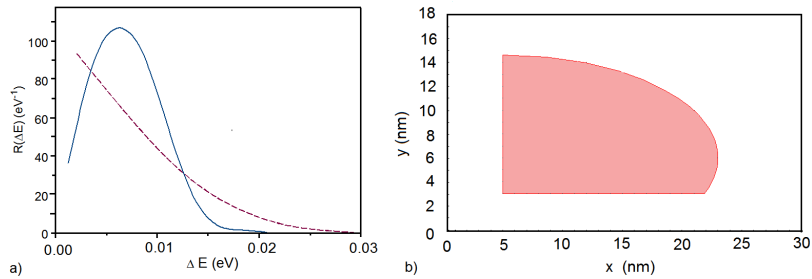

Figure 16. Violation of the shape Up-Down symmetry for two dimensional Si/SiO$_2$ QD. a) Distribution functions for electron neighboring levels for the "regular" semi-ellipsoidal shape (dashed curve), for the semi-ellipsoidal shape with the cut (solid curve). b) The shape of the QD with cut (in Cartesian coordinates).

Thus, we consider tunnel coupled two QDs with substrate between, which serves as barrier with electronic properties distinct from QD. Boundary conditions for the single electron Schrödinger equation are standard. We take into account the mass asymmetry inside as well outside of QDs [47]. To avoid the complications connected with spin-orbit coupling, $s$-levels of electron are only considered in the following. We would like to remind that the selection of levels with the same quantum numbers is requisite for study of NNS and other types of level statistics.

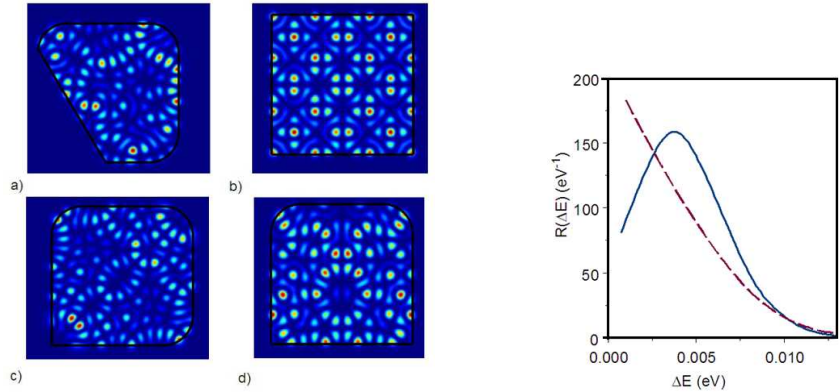

Figure 17. Shape of the 2D InAs/GaAs quantum dots (Left). The black curves mean the perimeters. The electron wave function contour plots of the excited state (with energy about 0.5 eV are shown). The corresponding types of the level statistics are shown (Right). The shape a) leads to non-Poissonian statistics (solid curve). The shapes b)-d) result to the Poissonian statistics (dashed curve).

Whereas at the large distances between dots each dot is independent and electron levels are twofold degenerate, expressing the fact that electron can be found either in one or in the other isolated dot, at the smaller inter-dot distances the single electron wave function begins to delocalize and extends to the whole DQD system. Each twofold degenerated level of the SQD splits by two, difference of energies is determined by the overlap, shift and transfer integrals [50]. Actually, due to the electron spin, there is fourfold degeneracy, however that does not change our results and below we consider electron as spinless. Note that the distance of removing degeneracy is different for different electron levels. This distance is larger for levels with higher energy measured relative to the bottom quantum well (see Fig. 21 below). By the proper choice of materials of dots and substrate one can amplify the "penetration" effects of the wave function.

Below we display some of our results for semiconductor DQDs. The band gap models are given in Ref. [47]. Fig. 18 shows distribution function for two $Si/SiO_2$ QDs of the shape of the 3D ellipsoids with a cut below the major axis. Isolated QD of this shape, as we saw in the previous section, is strongly chaotic. It means that distribution function of this QD can be well fitted by Brody formula with the parameter which is close to unity [47]. We see that the corresponding up-down mirror symmetric DQD shows Poisson-like NNS. Note that these statistics data involved 300 confined electron levels, which filled the quantum well from bottom to upper edges. We considered the electron levels with the orbital momentum $l=0$, as was mentioned above. The orbital momentum of electron can be defined due to rotational symmetry of the QD shape.

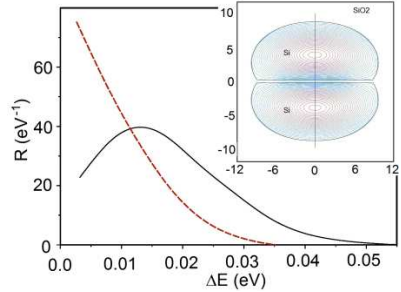

Figure 18. The electron wave function of the ground state is shown by the contour plot. (The lower figure) Distribution functions for energy differences of the electron neighboring levels in Si/SiO$_2$ single QD (solid line) and DQD (dashed line). The coefficient of the spline smoothing is equal 6. The cross section of DQD shape is shown in inset (sizes are given in nm).

In Fig. 19, SQD (2D quantum well) without both type of symmetry reveals level repulsion, two tunnel coupled dots show the level attraction. From the mirror symmetry point of view, the chaotic character of such single object is due to the lack of the R-L and up-down mirror symmetries. The symmetry requirements in this case, for the coupled dots are less restrictive: presence of one of the mirror symmetry types is sufficient for the absence of quantum chaos.

Dependence NNS on the interdot distance shows a gradual transition to the regular behaviour with intermediate situation when Poisson-like behavior coexists with chaotic one: they combine but the level attraction is not precisely Poisson-like. Further decreasing distance restores usual Poisson character (see Fig. 18). Fig. 19 shows how the degeneracy gradually disappears with the distance $b$ between QDs in InAs/GaAs DQD.

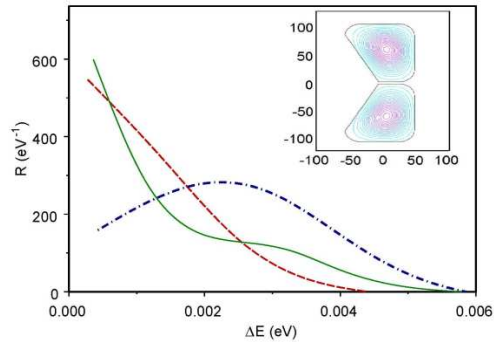

Figure 19. Distribution functions for energy differences of the electron neighboring levels in the 2D InAs/GaAs DQW calculated for various distances $b$ between QWs. Dashed (solid) line corresponds to $b$ =4 nm ($b$ =2 nm). Distribution functions of single QW is also shown by the dot-dashed line. The DQW shape is shown in inset (sizes are in nm).

Finally, we would like to show the disappearance of the Quantum Chaos when chaotic QW is involved in the "butterfly double dot" [46] giving huge conductance peak in the semi-classical approach. Fig. 20 shows the NNS for chaotic single QW of Ref. [46]

by dashed line. Mirror (up-down and L-R) symmetry is violated. The NNS for an L-R mirror symmetric DQW is displayed by solid line in Fig. 20. It is clear that Quantum Chaos disappears.

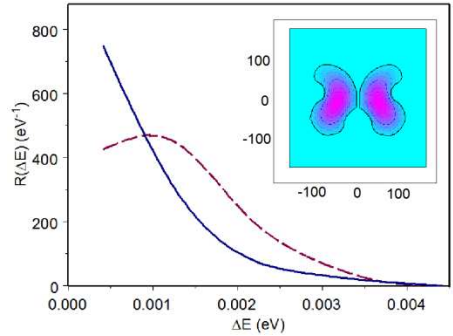

Figure 20 (24). Distribution functions for electron neighboring levels in InAs/GaAs single QW (dashed line) and DQW (solid line). Shape of DQW is shown in the inset. The electron wave function of the ground state is shown by the contour plot in the inset. Data of the statistics include 200 first electron levels.

We conjecture that the above mentioned peak in conductance of Ref. [46] and observed here a disappearance of Quantum Chaos in the same array are the expression of the two faces of the Quantum Mechanics with its semi-classics and genuine quantum problem of the energy levels of the confined objects, despite the different scales (what seems quite natural) in these two phenomena (several micrometers and 10–100 nm, wide barrier in the first case and narrow one in the second). We have to emphasize here that the transport properties are mainly the problem of the wave function whereas the NNS is mainly the problem of eigenvalues. Similar phenomena are expected for the several properly arranged coupled multiple QDs and QD superlattices. In the last case, having in mind, for simplicity, a linear array, arranging the tunnel coupling between QDs strong enough, we will have wide mini-bands containing sufficient amount of energy levels and the gap between successive mini-bands will be narrow. Since the levels in the different mini-bands are uncorrelated, the overall NNS will be Poissonian independently of the chaotic properties of single QD. We would like to remark also that our results have place for 3D as well as for 2D quantum objects. It is important to notice that the effect of reduction of the chaos in a system of DQD could appear for interdot distances larger than considered, for instance see Fig. 18, if an external electrical field is applied. By properly designed bias the electric field will amplify wave function "penetration" effectively reducing a barrier between QDs.

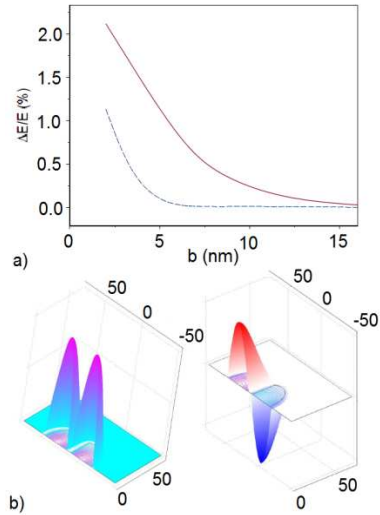

Figure 21. (The upper figure) Doublet splitting $\Delta E$ of single electron levels dependence on the distance $b$ between QDs in InAs/GaAs DQD. The ground state ( $E$ =0.23 eV) level splitting is expressed by dashed line. The solid line corresponds to doublet splitting of a level which is close to upper edge of the quantum well ( $E$ =0.56 eV). The shape of DQD is the same as in Fig. 18 (The lower figure). The electron wave functions of the doublet state: the ground state (left) and first excited state (right), are shown.

Thus, we have shown that the tunnel coupled chaotic QDs in the mirror symmetric arrangement have no quantum chaotic properties, NNS shows energy level attraction as should to be for regular, non-chaotic systems. These results are confronted with the huge conductance peak found by the semi-classical method in Ref. [46]. We think that our results have more general applicability for other confined quantum objects, not only for the quantum nanostructures, and may be technologically interesting. Concerning the last issue, problem is what easer: try to achieve regular, symmetric shape of SQDs, or, not paying attention to their irregular, chaotic shape arranges more or less symmetric mutual location [51].

### 9.2. Electron transfer between rings of Double Concentric Quantum Ring in magnetic field

Quantum rings are remarkable meso- and nanostructures due to their non-simply connected topology and attracted much attention last decade. This interest supported essentially by the progress in the fabrication of the structures with wide range of geometries including single and double rings. This interest rose tremendously in the connection to the problem of the persistent current in mesoscopic rings [52]. Transition from meso - to nano -scale makes more favorable the coherence conditions and permits to reduce the problem to the few or even to single electron.

Application of the transverse magnetic field $B$ leads to the novel effects: Whereas the quantum dots (QDs) of the corresponding shape (circular for two dimensional (2D), cylindrical or spherical for 3D ) has degeneracy in the radial $n$ and orbital $l$ quantum numbers, QR due to the double connectedness in the absence of the magnetic field $B$ has degeneracy only in $l$, and at the nonzero $B$ lifts the degeneracy in $l$, thus making possible the energy level crossing at some value of $B$, potentially providing the single electron transition from one state to the another.

Use the configurations with double concentric QR (DCQR) reveals a new pattern: one can observe the transition between different rings in the analogy with atomic phenomena. For the DCQR, the 3D treatment is especially important when one includes the inter-ring coupling due to the tunneling. The dependence on the geometries of the rings (size, shape and etc.) becomes essential.

We investigate the electron wave function localization in double concentric quantum rings (DCQRs) when a perpendicular magnetic field is applied. In weakly coupled DCQRs can be arisen the situation, when the single electron energy levels associated with different rings may be crossed. To avoid degeneracy, the anti-crossing of these levels has a place. In this DCQR the electron spatial transition between the rings occurs due to the electron level anti-crossing. The anti-crossing of the levels with different radial quantum numbers (and equal orbital quantum numbers) provides the conditions when the electron tunneling between rings becomes possible. To study electronic structure of the semiconductor DCQR, the single sub-band effective mass approach with energy dependence was used (see section 2). Realistic 3D geometry relevant to the experimental DCQR fabrication was employed taken from Refs. [53, 54]. The GaAs QRs and DQRs rings, embedded into the Al0.3Ga0.7As substrate, are considered [55]. The strain effect between the QR and the substrate materials was ignored here because the lattice mismatch between the rings and the substrate is small. Due to the non-parabolic effect taken into account by energy dependence effective mass of electron in QR, the effective mass of the electron ground state is calculated to be the value of $0.074\,m_0$ that is larger than the bulk value of $0.067\,m_0$. For the excited states, the effective mass will increase to the bulk value of the Al0.3Ga0.7As substrate. Details of this calculation one can find in Ref. [55].

Electron transfer in the DCQR considered is induced by external factor like a magnetic or electric fields. Probability for this transfer strongly depends on the geometry of DCQR. The geometry has to allow the existing the weakly coupled electron states. To explain it, we note that DCQR can be described as a system of double quantum well. It means that there is duplication of two sub-bands of energy spectrum (see [9] for instance) relative the one for single quantum object. In the case of non-interacting wells (no electron tunneling between wells) the each sub-band is related with left or right quantum well. The wave function of the electron is localized in the left or right quantum well. When the tunneling is possible (strong coupling state of the system), the wave function is spread out over whole volume of the system. In a magnetic field, it is allowed an intermediate situation (weak coupled states) when the tunneling is possible due to anti-crossing of the levels. Anti-crossing, of course, is consequence of the impossibility to cross of levels with the same space symmetry [56, 57].

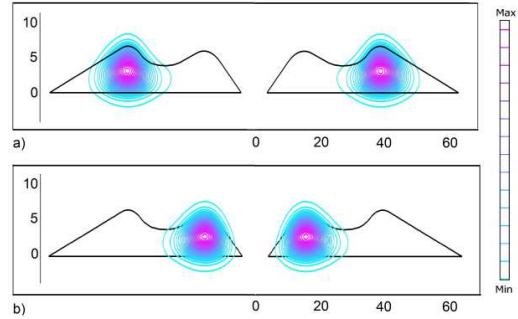

Figure 22. The squares of wave functions for the a) $(1,0)$, outer ($E = 0.072$ eV) and b) $(2,0)$, inner ($E = 0.080$ eV) states are shown by contour plots. The contour of the DCQR cross-section is given. The sizes are in nm.

There is a problem of notation for states for DCQR. If we consider single QR (SQR) then for each value of the orbital quantum number $|l| = 0,1,2...$ in Eq. (2) we can definite radial quantum number $n = 1,2,3,...$ corresponding to the numbers of the eigenvalues of the problem (7) in order of increasing. One can organize the spectrum by sub-bands defined by different $n$. When we consider the weakly coupled DCQR, in contrast of SQR, the number of these sub-bands is doubled due to the splitting the spectrum of double quantum object [50]. Electron in the weakly coupled DCQR can be localized in the inner or outer ring. In principle, in this two ring problem one should introduce a pair of separate sets of quantum numbers $(n_i, l)$ where index $i = 1,2$ denoted the rings where electron is localized. However, it is more convenient, due to the symmetry of the problem, to have one pair $(n,l)$ numbers ascribed to both rings (inner or outer), in other words, we use a set of quantum numbers $(n,l), p$ where $p$ is dichotomic parameter attributed to the electron localization ("inner" or "outer").

Since we are interested here in the electron transition between rings and, as we will see below, this transition can occur due to the electron levels anti-crossing followed a tunneling, we concentrate on the changing of the quantum numbers $n$. The orbital quantum numbers must be equal providing the anti-crossing of the levels with the same symmetry (see [57]). Thus, the anti-crossing is accompanied by changing the quantum numbers $n$ and $p$ of the $(n,l), p$ set.

Strongly localized states exist in the DCQR with the geometry motivated by the fabricated DCQR in [53, 54]. The wave functions of the two $s$-states of the single electron with $n = 1,2$ are shown in Fig. 22, where the electron state $n = 1$ is localized in outer ring, and the electron state $n = 2$ is localized in inner ring. Moreover all states of the sub-bands with $n = 1,2$, and $|l| = 1,2,3...$ are well localized in the DCQR. The electron localization is outer ring for $n = 1$, $|l| = 0,1,2,...$, and inner ring for $n = 2$, $|l| = 0,1,2....$

The difference between spectra of the two sub-bands can be explained by competition of two terms of the Hamiltonian of Eq. (7) and geometry factor. The first term includes first derivative of wave function over $\rho$ in kinetic energy; the second is the centrifugal term. For $|l| \neq 0$ the centrifugal force pushes the electron into outer ring.

One can see that the density of the levels is higher in the outer ring. Obviously, the geometry plays a role also. In particular, one can regulate density of levels of the rings by changing a ratio of the lateral sizes of the rings.

Summarizing, one can say that for $B=0$ the well separated states are only the states $(1,l), p$ and $(2,l), p$. Thus, used notation is proper only for these states. The wave functions of the rest states $(n>2,l)$ are distributed between inner and outer rings. These states are strongly coupled states.

Crossing of electron levels in the magnetic field $B$ are presented in Fig. 23 There are crossings of the levels without electron transfer between the rings. This situation is like when we have crossing levels of two independent rings. There are two crossings when the orbital quantum number of the lower state is changed due to the Aharonov-Bohm effect. It occurs at about 0.42 T and 2.5 T. There are two anti-crossings: the first is at 4.8 T, another is at 5.2 T. These anti-crossings are for the states with different $n$; the first are states (1,0) and (2,0) and the second are states (1,-1) and (2,-1). In these anti-crossings the possibility for electron tunneling between rings are realized. In Fig. 24 we show how the root mean square (rms) of the electron radius is changed due to the tunneling at anti-crossing. One can see from Fig. 23 that the electron transition between rings is only possible when the anti-crossed levels have different radial quantum numbers and equal orbital quantum numbers, in accordance of Ref. [56].

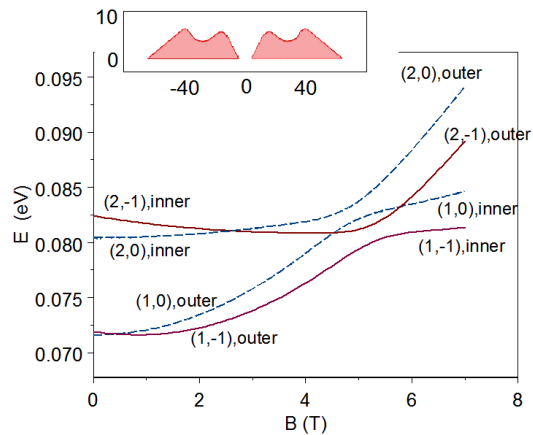

Figure 23. Single electron energies of DCQR as a function of magnetic field magnitude $B$. Notation for the curves: the double dashed (solid) lines mean states with $l=0$ ($l=-1$) with $n=1,2$. The quantum numbers of the states and positions of the electron in DCQR are shown. The cross section of the DCQR is given in the inset.

Transformation of the profile of the electron wave function during the process of anti-crossing with increasing $B$ is given in Fig. 25. The electron state (1,-1), outer is considered as "initial" state of an electron ($B=0$). The electron is localized in outer ring. Rms radius is calculated to be $R=39.6$ nm. For $B=5.2$ T the second state is the tunneling state corresponding to the anti-crossing with the state (0,-1). The wave function is spreaded out in both rings with $R=32.7$ nm. The parameter $p$ has no definite value for this state. The "final" state is considered at $B=7$ T. In this state the electron was localized in inner ring with $R=17.6$ nm. Consequently connecting these three states of the electron, we come to an electron trapping, when the electron of outer ring ("initial" state) is transferred to the inner ring ("final" state). The transfer process is governed by the magnetic field.

Note that the energy gap between anti-crossed levels which one can see in Fig. 26 can be explained by the general theory for double interacting quantum well [50]. The value of the gap depends on separation distance between the rings, governed by the overlapping wave functions corresponding to the each ring, and their spatial spread which mainly depends on radial quantum number of the states [55].

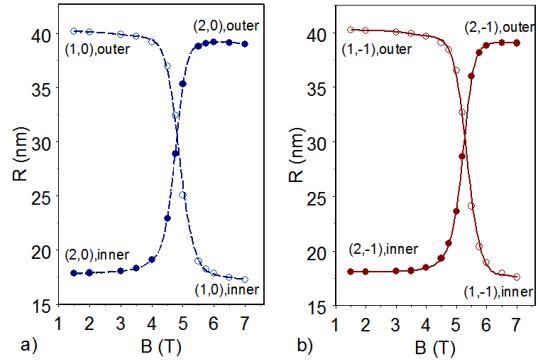

Figure 24. Rms radius of an electron in DCQR as a function of magnetic field for the states a) $((n=1,2), l=0)$ and b) $((n=1,2), l=-1)$ near point of the anti-crossing. The calculated values are shown by solid and open circles. The dashed (solid) line, associated with states of $l=0$ ($l=-1$), fits the calculated points.

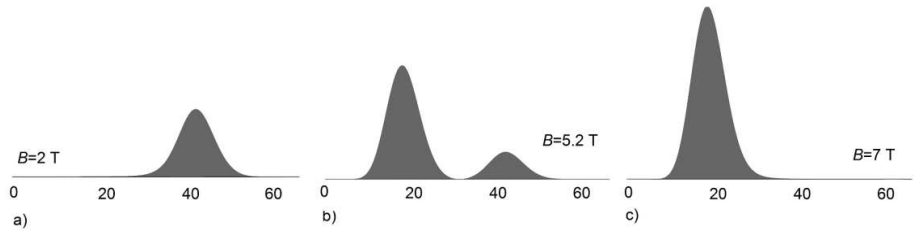

Figure 25. Profiles of the normalized square wave function of electron in the states a) $(1,-1)$,outer; b) $(1,-1)$,n/a and c) $(1,-1)$,inner for different magnetic field $B$. The a) is the "initial" state ($B=0$) with $R=39.6$ nm, the b) is the state of electron transfer ($B=5.2$ T) with $R=32.7$ nm, the c) is the "final" state ($B=7$ T) with $R=17.6$ nm. The radial coordinate $\rho$ is given in nm (see Fig. 22 for the DCQR cross section).

Other interesting quantum system is one representing QR with QD located in center of QR. The cross section of such heterostructure (GaAs/Al0.3Ga0.7As) is shown in Fig.

26a. In Fig. 26b we present the results of calculations for electron energies of the (1,0) and (3,0) states in the magnetic field $B$ [55]. Once more we can the level anti-crossing (for about of 12.5 T). This anti-crossing is accompanied by exchange of electron localization between the QD and the QR. It means that if initial state (for $B$ <12.5 T) of electron was the state (1,0),outer, then the "final" state (for $B$ >12.5 T) will be (1,0),inner. It can be considered as one of possibilities for trapping of electron in QD.

One can see from Fig. 26b that the energy of the dot-localized state grows more slowly than the envelope ring-localized state. At the enough large $B$ the dot-localized state becomes the ground state [58]. In other words, when the Landau orbit of electron becomes smaller then dot size, electron can enter the dot without an extra increase of kinetic energy.

Concluding, we made visible main properties of this weakly coupled DCQD established by several level anti-crossings that occurred for the states with different radial quantum number $n$ ($n$ =1,2) and equal orbital quantum number $l$. One may conclude that the fate of the single electron in DCQRs is governed by the structure of the energy levels with their crossing and anti-crossing and changing with magnetic field. The above described behavior is the result of the nontrivial excitation characteristic of the DCQRs. Effect of the trapping of electron in inner QR (or QD) of DCQR may be interesting from the point of view of quantum computing.

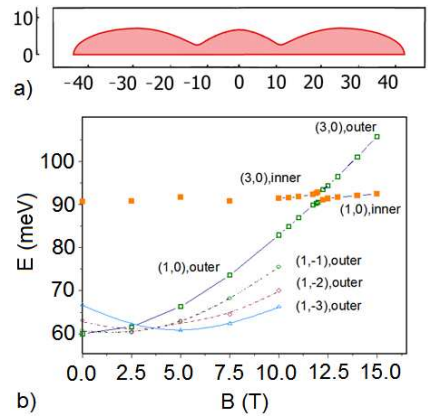

Figure 26. a) Cross section of the QR with QD system. Sizes are given in nm. b) Energies of the (1,0) and (3,0) states in the magnetic field $B$ for the QR with QD system. The open symbols show that the electron is localized in the ring. The solid squares show that the electron localized in QD.

**Acknowledgment**

This work is supported by NSF CREST award; HRD-0833184 and NASA award NNX09AV07A.